\documentclass[reprint,
superscriptaddress,
 amsmath,amssymb,
 aps,
pra,
]{revtex4-2}

\usepackage{graphicx}
\usepackage{dcolumn}
\usepackage{bm}
\usepackage{physics}
\usepackage{ulem}
\usepackage{hyperref}
\hypersetup{
	colorlinks = true,
	urlcolor   = blue,
	linkcolor  = blue,
	citecolor  = red
}
\usepackage{amsmath}
\usepackage{algorithm}
\usepackage{algpseudocode}
\usepackage{float}
\usepackage{siunitx}
\usepackage{comment}

\begin{document}

\title{ Controlling Sloppiness in Two-Phase Estimation with a Tunable Weak Measurement}

\author{G. Bizzarri}
\affiliation{Dipartimento di Scienze, Universit\`a degli Studi Roma Tre, Via della Vasca Navale, 84, 00146 Rome, Italy}

\author{M. Parisi}
\affiliation{Dipartimento di Scienze, Universit\`a degli Studi Roma Tre, Via della Vasca Navale, 84, 00146 Rome, Italy}

\author{M. Manrique}
\affiliation{Dipartimento di Scienze, Universit\`a degli Studi Roma Tre, Via della Vasca Navale, 84, 00146 Rome, Italy}

\author{I. Gianani}
\affiliation{Dipartimento di Scienze, Universit\`a degli Studi Roma Tre, Via della Vasca Navale, 84, 00146 Rome, Italy}

\author{A. Chiuri}
\affiliation{Nuclear Department, ENEA, Via E. Fermi 45, 00100 Frascati, Italy}

\author{M. Rosati}
\affiliation{Dipartimento di Ingegneria Civile, Informatica e delle Tecnologie Aeronautiche, Universit\`a degli Studi Roma Tre, Via Vito Volterra 62, 00146 Rome, Italy}

\author{V. Giovannetti} 
\affiliation{Scuola Normale Superiore, 56126 Pisa, Italy}

\author{M.G.A Paris}
\affiliation{Dipartimento di Fisica ``Aldo Pontremoli", Universit\`a degli Studi di Milano, Via Celoria, 16, 20133 Milan, Italy}

\author{M. Barbieri}
\affiliation{Dipartimento di Scienze, Universit\`a degli Studi Roma Tre, Via della Vasca Navale, 84, 00146 Rome, Italy}
\affiliation{Istituto Nazionale di Ottica - CNR, Largo E. Fermi 6, 50125 Florence, Italy}

\begin{abstract}
The description of complex systems requires a progressively larger number of parameters. However, in practice, it often happens that a small subset of parameters suffices to describe the dynamics of the system itself: \color{black}{these combinations are usually referred to as \textit{stiff} combinations. In turn, the remaining combinations, called \textit{sloppy}, only play a minor role in the dynamics of the system, hence provide little information on it}. While this effect can reduce model complexity, it can also limit the estimation precision when the stiff and sloppy combinations are unknown to the experimenter, and one is forced to estimate the potentially sloppy model parameters. We explored how such a sloppy behavior can be controlled and counteracted via quantum weak measurements in the estimation of two sequential phases. We showed that the introduction of a weak measurement of variable strength in-between the two phases allows to switch from a fully sloppy setup to a fully determined one where both phases can be estimated with quantum-limited precision.
Our work provides an important insight of sloppiness detection in quantum systems, with promising applications in quantum metrology and imaging, as well as to quantum security and quantum monitoring.
\end{abstract}
\maketitle
\section{Introduction}
Understanding complex systems often demands the availability of a model to guide us through the different aspects of their behaviour. By comparing observations and predictions, one can then infer the value of the relevant parameters appearing in the model and gain further predictive power. As  models are made more refined, the number of necessary parameters typically grows. Nevertheless, the actual observations may be dictated by only a small combination of said parameters. For instance, we can include  temperature, pressure, concentration, etc., in the list of parameters influencing biological systems, but, in practice, distinct arrangements of their values may lead to identical behaviours, due to the very nature of the phenomenon or because of some active reaction mechanism. Such an occurrence of a model with many parameters being actually governed by a lesser number of combined parameters is called 'sloppiness' ~\cite{brown2003statistical,brown2004statistical}, and it is frequently encountered, especially in systems of interest for biology~\cite{PhysRevLett.97.150601,machta2013parameter}. 

\color{black}{The evolution of a sloppy model is thus governed by such few \textit{stiff} combinations, as they are commonly indicated in the field of biological complex systems. Consequently, observing the dynamics may provide substantial information on their values. In quantitative terms, this means that measurements can be performed with high Fisher information on those parameters. However, such stiff combinations coexist with other \textit{sloppy} combinations (once again, borrowing from the same terminology), that are, conversely, hardly relevant in determining the system's behaviour. From the point of view of metrology, no measurement applied to the system would show significant Fisher information. In plain terms, stiff parameters can be retrieved with sufficient precision, whereas sloppy parameters are bound to be loosely determined.}

On the one hand, the presence of stiff parameters preserves a model's robustness to perturbations and enables the identification of key physical quantities determining the observed behaviour. This is particularly relevant in complex quantum systems characterized by a massive number of parameters, where quantum metrology generally prohibits the simultaneous estimation of multiple parameters with maximum precision; if a few stiff parameters are present, one can focus on estimating them with better precision. On the other hand, if the stiff and sloppy combinations are unknown, one should attempt estimating the initial parameters, but it entails an unavoidable loss of precision. 
A partial remedy to such an issue is found by altering in a controlled fashion the state of the system, so that it can evolve according to the new, perturbed conditions, rather than the standard ones. This intervention can thus lead to an increase of the available information on the model parameters, and a consequent reduction of its sloppiness. However, there could be instances in which it is advisable to keep such modifications to a small extent, in order to keep the dynamics close to the natural case.

\begin{figure}[h!]
    \centering
    \includegraphics[width=\columnwidth]{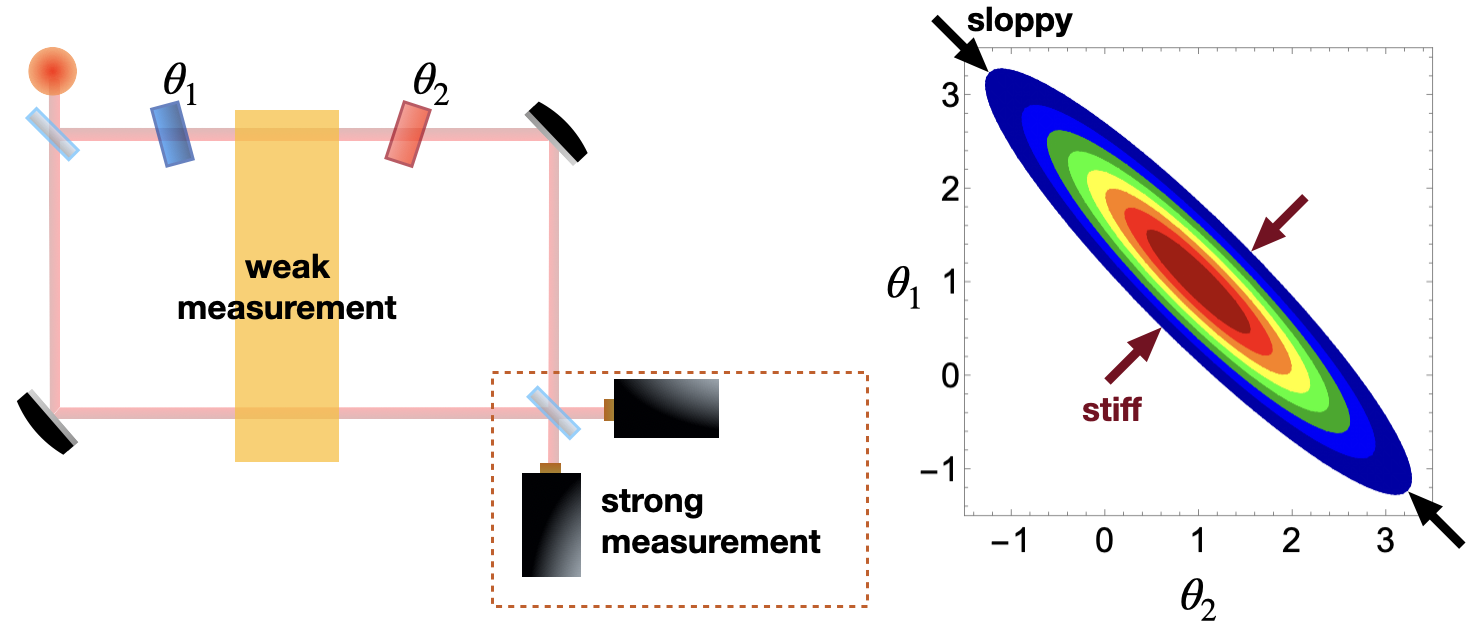}
    \caption{Scheme of a {\color{black} generic} sloppy two-phase estimation. Left: light in an interferometer acquires  phase shifts $\theta_1$ and $\theta_2$ from two consecutive objects. A weak measurement is carried out in between in order to disambiguate the two phases. 
Right:  the two values of $\theta_1$ and $\theta_2$ can now be retrieved, with a joint uncertainty qualitatively represented as an ellipse.
The parameters appear in a well determined \textit{stiff} combination $\theta_1+\theta_2$  and, crucially, in a poorly determined \textit{sloppy} one. }  
    \label{fig:scheme}
\end{figure}

Sloppiness \color{black}{has been recognised to} emerge in quantum models as well~\cite{PRXQuantum.2.020308,Goldberg21,Yang23,Frigerio2024,PhysRevA.111.012414}, due to the same asymmetry in the available information of stiff and sloppy combinations, with the added interplay of quantum incompatibility in parameter estimation. Its scrutiny is engaging not only for answering fundamental questions, but also for designing new protocols for secure quantum sensing. Indeed, sloppiness can be actively pursued as a way of performing distributed sensing in a secure way~\cite{bugalho2024private,hassani2024privacy,ho2024quantum,Bizzarri2025}. This makes it possible, for instance, to infer a specific combination of local parameters in a network while keeping each individual parameter secret.  A relevant exemplifying case is presented in Fig.~\ref{fig:scheme}: two phases $\theta_1$ and $\theta_2$ are accumulated in two distinct processes on one arm of a Mach-Zehnder interferometer (MZI). Any detector would then produce results only depending on the sum $\theta_1+\theta_2$, the stiff parameter of the system, while the difference $\theta_1-\theta_2$ is completely undetermined. \color{black}{If one is interested in the unspoiled dynamics of the system, the knowledge of the stiff parameter $\theta_1+\theta_2$ would be sufficient for its characterisation. However, sloppiness makes it impossible to retrieve the values of the individual phases $\theta_1$ and $\theta_2$, whenver these are associated to processes of interest.}

Reducing ambiguity through perturbation represents a radical departure from the conventional three-act framework of quantum metrology: state preparation, evolution, and measurement~\cite{Giovannetti2006,Paris2009}. Traditionally, the evolution step is considered fixed and unmodifiable by the experimentalist, whereas now, an intervention is necessary to eliminate sloppiness. \color{black}{We can draw inspiration from control theory in order to conceive and categorise such strategies. The first option, akin to open-loop schemes, contemplates introducing an additional unitary operation in the midst of the evolution. This choice has been analysed by Yang et al. in the context of two-photon interferometry~\cite{Yang23}, with the aim of untwining the imprinting of parameters that would normally be associated to the same generators. This has been extended by Frigerio and Paris~\cite{Frigerio2024}, who have shown that scrambling may as well be used, without compromising favourable scaling in the precision.} 

\color{black}{The second option, instead, builds on close-loop scheme thus demanding to observe the system, possibly by means of the interaction with an auxiliary system. In this case, however,} access to the system may be limited, particularly since the interaction time must be kept short relative to other timescales in the evolution. This constraint, along with other potential physical limitations on the interaction and the goal of minimizing perturbation to the natural model, may result in an incomplete measurement of the system. In this article we propose the use of a weak measurement~\cite{Dressel2014a} to tackle model sloppiness with a non-invasive procedure. {\color{black} In our scheme, Fig.~\ref{fig:scheme}}, a weak measurement is introduced in-between the action of the phases $\theta_1$ and $\theta_2$, whose estimation constitutes a sloppy model in terms of a two-level quantum system, {\it i.e.} a qubit. 
Coding is then performed on the polarisation of a single photon and, thanks to the possibility of measuring it without necessarily destroying it~\cite{PhysRevLett.92.190402}, we investigate the consequences of extracting limited information for a two-phase estimation experiment. Our study demonstrates that the degree of sloppiness of the model can be controlled via the strength of the weak measurement, switching continuously from a fully sloppy setting to a perfectly determined one where both phases can be estimated with quantum-limited precision.



\section{Results}

\subsection {Two-phase sloppy estimation} \color{black}{In our scheme, we implement the two arms of the MZI as two orthogonal polarisations of a single spatial mode, as this is key to maintaining phase stability. A single photon is thus prepared in an equal superposition of the right-circular $\vert R\rangle$ and the left-circular $\vert L \rangle$ orientation,} {\it viz.} $\vert H\rangle = \left(\vert R\rangle+\vert L\rangle\right)/\sqrt{2}$ ($H$ and $V$ stand for the horizontal and vertical polarisation, respectively, with $\ket V =  \left(\vert R\rangle+\vert L\rangle\right)/(\sqrt{2}\text{i})$). The problem can then be described as the evolution of a qubit, in particular, the action of each phase-shifter is a rotation $U(\theta)=e^{-2i\theta Y}$. Here  $Y$ is the $y$ Pauli operator in the basis $\{\ket{H},\ket{V}\}$, equivalent to the photon-number unbalance between the two modes.  Consequently, the unperturbed evolution of the state yields the output state $U(\theta_2)U(\theta_1)\vert H\rangle=\cos 2(\theta_1+\theta_2)\vert H\rangle+\sin 2(\theta_1+\theta_2)\vert V\rangle$, in which the action of the individual elements cannot be isolated. We notice that, in formal terms, the two transformations $U(\theta_1)$ and $U(\theta_2)$ share the same generator, stressing that sloppiness is a problem of classical statistics, and its origin should not be traced back to aspects like non-commutativity. Nevertheless, these may have an interplay with sloppiness in the presence of multiple parameters.

In order to reduce sloppiness, this evolution is modified by inserting a weak measurement in-between the two phase-shifters. Clearly, in order to have an effect on sloppiness, this ought to be sensitive to the coherence between the two modes, otherwise it would extract no information on the phase. Therefore, the weak measurement should be associated to an unbiased observable with respect to $Y$, for instance, to the Pauli $Z$ observable of the qubit, corresponding to discriminating the $H$ and the $V$ polarisations. This is realised by coupling the system qubit to a second meter qubit, and then measuring the latter. It is well-known that, due to the correlations established by the coupling, this operation provides information about the original system. There is no guarantee, however, that their interaction is sufficiently strong to extract complete information, which would correspond to a fully projective measurement. 

A commonplace model for this measurement scheme employs a logic gate, {e.g.} a control-$Z$ gate $C_Z$ in the basis $\{\ket{H},\ket{V}\}$, as a template for the interaction. It takes as its inputs the system qubit after the first phase-shifter,  $U(\theta_1)\vert H\rangle$, and the meter in a generic state
$\ket{\mu}=\kappa \ket{D}+\sqrt{1-\kappa^2}\ket{A}$ (with $\ket{D}=(\ket{H}+\ket{V})/\sqrt{2}$, $\ket{A}=(\ket{H}-\ket{V})/\sqrt{2}$). When $\kappa=1$, the output of this gate is a maximally entangled state: a measurement of the meter in the $D/A$ basis corresponds to measuring $Z$ on the system in that it leads to the same probability and wavepacket reduction as a direct measurement of this observable on the input of the system. For $\kappa=1/\sqrt{2}$, the two-qubit state remains separable, thus a measurement on the meter can give no information on the system. In the intermediate cases, the coupling delivers a weak version of a $Z$ measurement, with the coefficient $K=2\kappa^2-1$ quantifying the amount of information available~\cite{PhysRevLett.92.190402}. Following the measurement, the system goes through the second phase element and is finally measured with an ordinary projective measurement of $Z$.  The overall measurement protocol has thus four outcomes, two for both the intermediate weak measurement and for the final strong measurement.

The metrological capabilities of the scheme are captured by the two-parameter Cram\'er Rao bound~\cite{Paris2009,albarelli2020perspective}, based on the Fisher information matrix $F$ associated with a measurement scheme. Its elements are given by $F_{jk}=\sum_x \left(\partial_{\theta_j}p(x|\theta_1,\theta_2)\partial_{\theta_k}p(x|\theta_1,\theta_2)\right)^2/p(x|\theta_1,\theta_2)$, where the index $x$ runs over the four possible outcomes, and the $p(x|\theta_1,\theta_2)$ are the corresponding measurement probabilities. Their knowledge allows us to infer the values of $\theta_1$ and $\theta_2$ repeating the measurement $N$ times, using an estimator linking the outcomes to the parameters. This bounds the covariance matrix $\Sigma$ of the estimators of $\theta_1$ and $\theta_2$ as $\Sigma \geq F^{-1}/N$, where $N$ is the number of events being recorded. This implies that the individual variances satisfy $\sigma^2_{\theta_k} \geq \left(F^{-1}\right)_{kk}/N$. 

\color{black}{This classical Fisher information matrix is subject to a matrix quantum Cram\'er-Rao bound $Q\geq F$ invoking the quantum Fisher information matrix (QFIM) $Q$. This only depends only on the joint state $\rho$ of the original qubit and the meter at the output of the gate: $\rho=\ket{\Psi}\bra{\Psi}$, with $\ket{\Psi}=(U(\theta_2)\otimes I)C_Z(U(\theta_1)\otimes I)\ket{H}\ket{\mu}$. It is defined as $Q_{jk}=\text{Tr}[\rho\left(L_{\theta_j}L_{\theta_k}+L_{\theta_k}L_{\theta_j}\right)]/2$, where the symmetric logarithmic derivatives $L_{\theta_j}$ are implicitly given by $\partial_{\theta_j}\rho= L_{\theta_j}\rho+\rho L_{\theta_j}$. An explicit calculation of the QFIM yields
\begin{equation}
\label{eq:fim}    
    Q = 16\begin{pmatrix}
        1&&\sqrt{1-K^2}\\
        \sqrt{1-K^2}&&1
    \end{pmatrix}.
\end{equation}
Our strategy of measurements on the system and meter qubits saturates the quantum Cram\'er-Rao bound $F=Q$, hence $\Sigma \geq Q^{-1}/N$. We remark that, in the general case, there is no insurance that a measurement scheme exists yielding to a saturation, since specific conditions must be met~\cite{Pezze2017}.} \color{black}{In addition, this establishes that a final strong measurement on a fixed observable $Z$ is sufficient, hence there is no need to implement a feedforward loop.}

The single-parameter quantum Cram\'er-Rao bound is the same for both phases and is given by
\begin{equation}
    \sigma^2_\theta\geq\frac{1}{16NK^2}.
\end{equation}
Notably, a fully projective measurement ($K = 1$) would yield the same information as measuring $\theta_1$ and $\theta_2$ in separate setups, whereas removing the intermediate measurement ($K=0$) leads to a fully singular quantum Fisher information matrix, preventing the estimation of the two phases. In addition, tuning the strength $K$ affects the \textit{correlation} between the two parameters. \color{black}{When a target variance $\bar\sigma^2$ is sought, the use of a weak measurement of strength $K$ entails a growth of the necessary resources by at least a factor $K^{-2}$ with respect the projective case.}

\color{black}{The measurement scheme imposes no trade-off on the individual precisions of $\theta_1$ and $\theta_2$. This is due to the fact the measurement strength $K$ sets the amount of information extracted from the first measurement - necessarily associated with $\theta_1$ - as well as the amount of coherence made available for the final projective measurement. Such a behaviour is reminiscent of other multiparameter schemes relying on entanglement with an ancilla~\cite{PhysRevA.97.032305,PhysRevA.97.042112,sciadv.abd2986,PhysRevLett.126.070503}, although these may actually require a joint measurement.}


We thus observe that the single-parameter precision obtained from the second (strong) measurement can be tuned by varying the strength of the first measurement.

The diagonalization of $F$ reveals that the sum $\theta_1 + \theta_2$ remains the stiff parameter in the model, with associated information $ F_+ = 16(1 + \sqrt{1 - K^2})$, while the difference $ \theta_1 - \theta_2 $ is the sloppy parameter, with associated information $F_- = 16(1 - \sqrt{1 - K^2})$. The sloppiness of the model is well captured by the determinant of $F = F_+ F_-=(16K)^2$~\cite{brown2004statistical}.  


\begin{figure}[h!]
    \centering
    \includegraphics[width=0.65 \columnwidth, trim={1cm .3cm .3cm .5cm},clip]{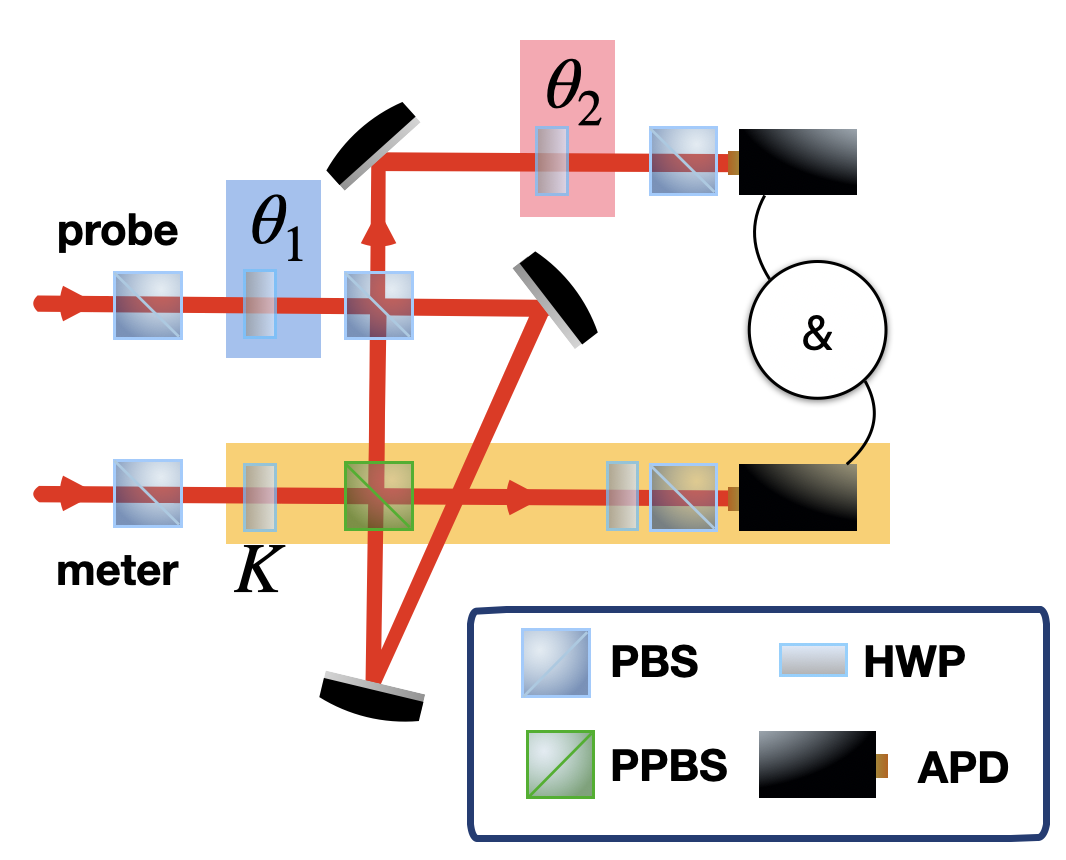}
    \caption{The experiment. Photonic qubits at central wavelength $\lambda_0 = 800 \si{\nano\meter}$ with polarization encoding are generated via a Spontaneous Parametric Down-Conversion (SPDC) source made of a $3\si{\milli\meter}$-thick $\beta$-Barium Borate (BBO) crystal pumped by a continuous-wave (CW) laser with typical power $P = 50 \si{\milli\watt}$. The two photons, representing the probe and meter qubit in our setup, are then sent to a control-Z gate embedded in a Sagnac loop. The logic gate exploits a partially-polarizing beam splitter (PPBS) featuring different transmittivities for horizontal and vertical polarizations, namely $T_H = 2/3$ and $T_V = 1/3$. Target phases $\theta_1$ and $\theta_2$, the measurement's strength $K$ and the Hadamard gate, mapping $\ket{H/V}$ to $\ket{D/A}$, are imparted by the rotation of half-waveplates (HWPs) . Photons are then collected by fiber-coupled avalanche photo-diodes (APDs).} 
    \label{fig:experiment}
\end{figure}

\subsection {Experimental results}\label{sec:exp_results} In order to test our predictions, we have performed an experiment with photon pairs, using one particle as the system and the other as the meter, see Fig.~\ref{fig:experiment}. The system is first initialised in the $\ket{H}$ state and then traverses the first phase element, a half-waveplate set an angle $\theta_1$. The {\rm C-Z} gate is the two-photon device previously demonstrated in~\cite{PhysRevLett.92.190402,ho2016experimental,bizzarri2024quasiprobability} , and requires to post-select events in which the two photons emerge on distinct arms. In our proof-of-principle demonstration we only consider these in the assessment of $N$. The strength $K$ of the weak measurement is conveniently set by the polarisation of the meter photon {\color{black} via a half wave-plate}~\footnote{The probabilistic behaviour of the gate implies that the input polarisation angles of both system and meter need being rescaled in order to account for extra loss in the vertical polarisation~\cite{PhysRevLett.95.210506}. }. Finally, a second half-waveplate at an angle $\theta_2$ works as the second phase-shifter.

\color{black}{Our estimation experiment proceeds in two steps. First, we obtain a calibration of our setup by collecting coincidence counts at different settings of $\theta_1$ (spanned from $0^\circ$ to $22.5^\circ$ with  uneven spacing), and $\theta_2$ (spanned in steps of $2.5^\circ$ from $0^\circ$ to $22.5^\circ$).  In particular, the measurements of the meter photon occur in the $D/A$ basis for all values of $K$, and in the $H/V$ basis for the probe photon. This calibration is collected at high statistics, in order to get a reliable estimate of the probability $p(x_w,x_s|\theta_1,\theta_2)$ for the outcome $x_s=H,V$ for the strong measurement, and  $x_w=D,A$ on the meter at the selected phase settings. These probabilities are then interpolated in between points by means of third-order polynomials, thus giving us access to their numerical value for arbitrary phases. This procedure has the advantage of naturally incorporating the genuine response of the setup, hence accounting for imperfections, but at the cost of not having an analytical expression available. The second step is the estimation experiment proper:  we have collected data corresponding to different pairs $(\theta_1,\theta_2)$, not included in the calibration set. The corresponding experimental counts $f(i,j)$ are used to derive a logarithmic maximum likelihood estimator as}
\begin{equation}
 (\bar \theta_1,\bar \theta_2)= \text{arg}\min_{\theta_1,\theta_2} \sum_{i,j}f(i,j) \ln{p(i,j|\theta_1,\theta_2)},
 \label{eq:maxlik}
\end{equation}
\color{black}{where the optimisation is run by numerical methods, hence does not require an explicit expression for the probability distribution.} The covariance matrix \color{black}{for each phase pair setting is obtained by a Monte Carlo method. This consists in repeating the estimation on coincidence counts, varied according to a Poisson distribution with average corresponding to the observed counts. The procedure goes under the name of bootstrapping.} The sample size in the estimation set is lower than for the calibration set (0.1 s $vs$ 5 s acquisition time per outcome combination). This reduces the impact of the precision on this reference~\cite{PhysRevLett.123.230502}.

\begin{figure*}[ht]
    \centering
    \includegraphics[width=\textwidth]{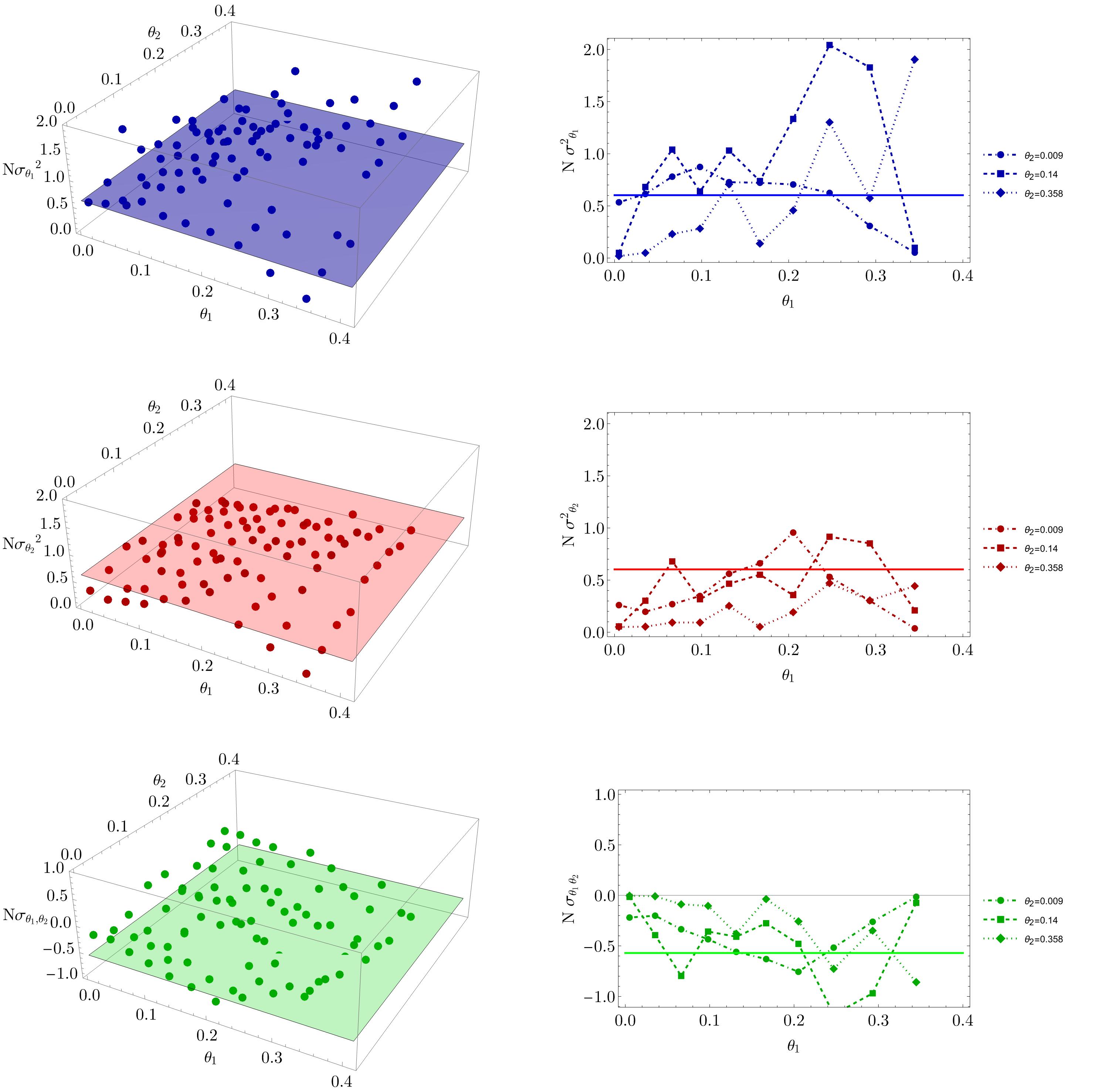}
    \caption{Elements of the covariance matrix $\Sigma$ for the two estimated phases $\theta_1$ and $\theta_2$ collected in the weak measurement condition $K=0.322$. From above, the data correspond to the variance on the 
    phase $\theta_1$ (top row), the variance on the phase $\theta_2$ (middle row), and their covariance (bottom row), all rescaled by the number $N$ of events collected. We show three-dimensional plots along with cuts at fixed values of $\theta_2$. In all plots, the points correspond to the experimental variances, and the solid curves to the predictions at the Cram\'er-Rao bound.} 
    \label{fig:Kweak}
\end{figure*}

\begin{figure*}[ht]
    \centering
    \includegraphics[width=\textwidth]{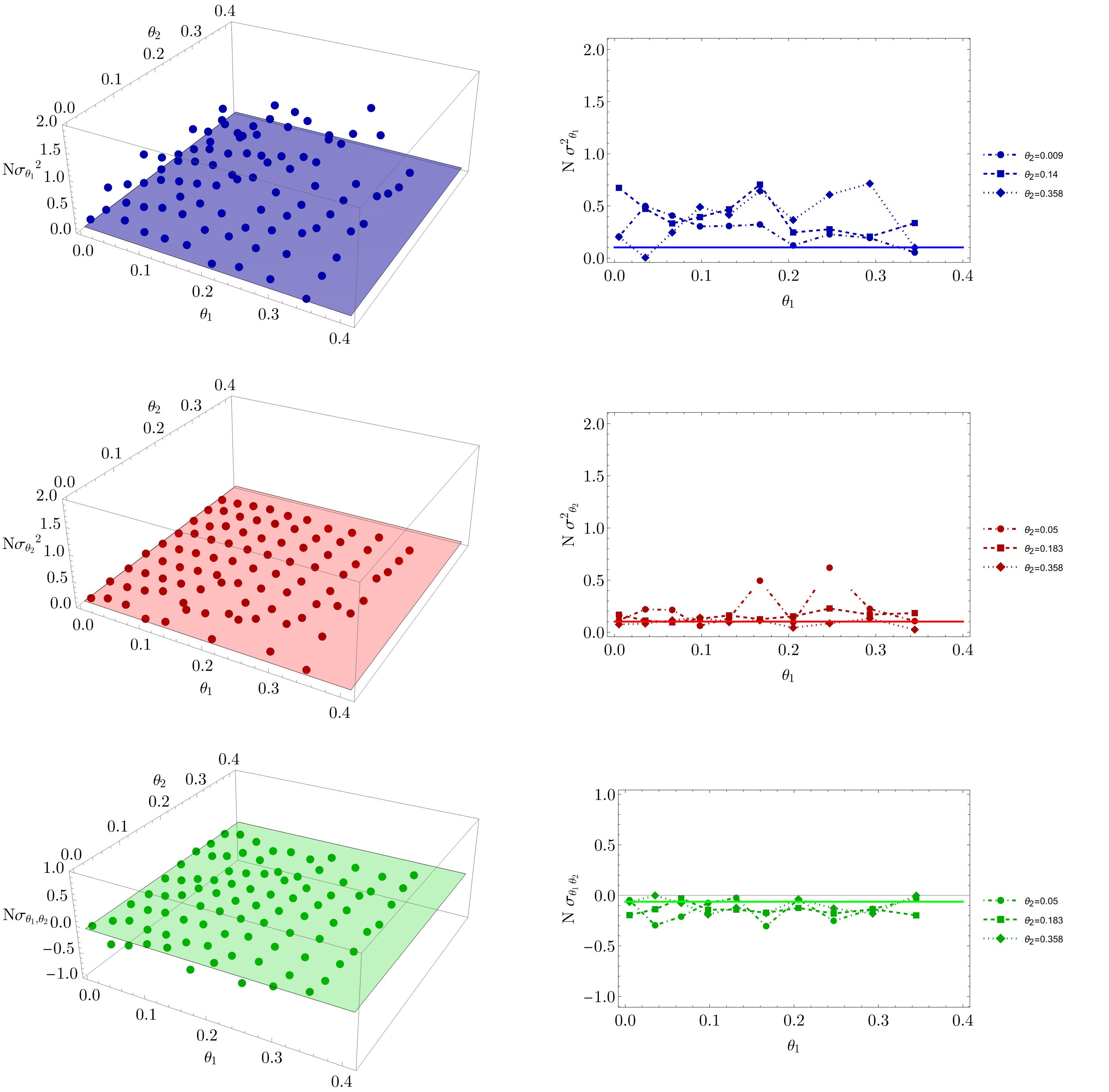}
    \caption{Elements of the covariance matrix $\Sigma$ for the two estimated phases $\theta_1$ and $\theta_2$ collected in the intermediate measurement strength condition $K=0.789$. From above, the data correspond to the variance on the phase $\theta_1$ (top row), the variance on the phase $\theta_2$ (middle row), and their covariance (bottom row), all rescaled by the number $N$ of events collected. We show three-dimensional plots along with cuts at fixed values of $\theta_2$. In all plots, the points correspond to the experimental variances, and the solid curves to the predictions at the Cram\'er-Rao bound.} 
    \label{fig:Kmedium}
\end{figure*}

\begin{figure*}[ht]
    \centering
    \includegraphics[width=\textwidth]{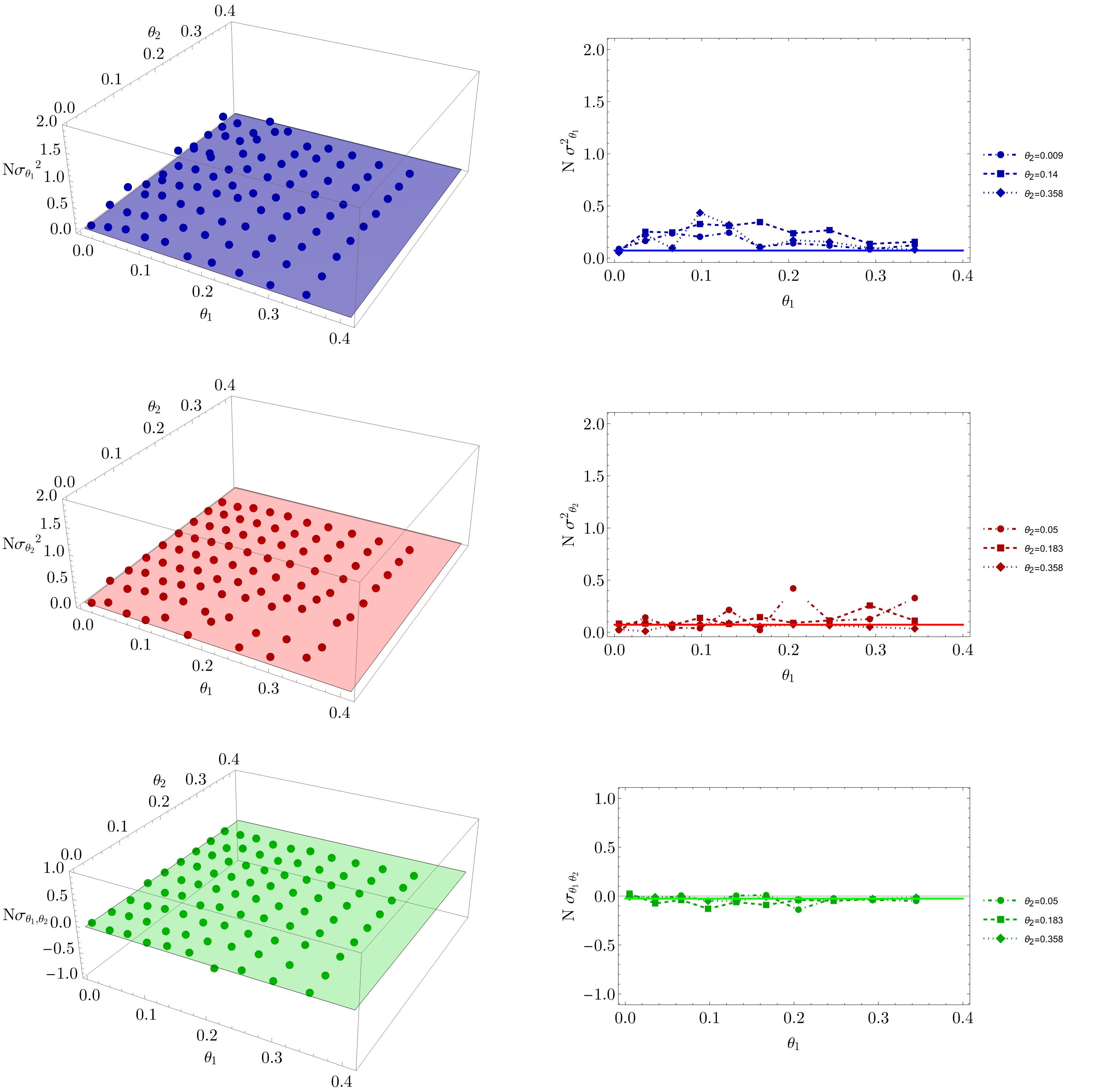}
    \caption{Elements of the covariance matrix $\Sigma$ for the two estimated phases $\theta_1$ and $\theta_2$ collected in a condition close to a projective measurement, $K=0.934$. From above, the data correspond to the variance on the phase $\theta_1$ (top row), the variance on the phase $\theta_2$ (middle row), and their covariance (bottom row), all rescaled by the number $N$ of events collected. We show three-dimensional plots along with cuts at fixed values of $\theta_2$. In all plots, the points correspond to the experimental variances, and the solid curves to the predictions at the Cram\'er-Rao bound.} 
    \label{fig:Kstrong}
\end{figure*}

\color{black}{The result on the achieved precision are reported in Fig.~\ref{fig:Kweak} for a weak condition, $K=0.322$, in Fig.~\ref{fig:Kmedium} for an intermediate strength, $K=0.785$, and, finally in Fig.~\ref{fig:Kstrong} for measurement close to a standard projector, $K=0.934$. In the weak condition, there are evident deviations from the ideal behaviour at the Cram\'er-Rao bound. These are attributed to the experimental imperfections, mostly imperfect nonclassical interference between the two photons. Indeed, in such conditions, the measurement can extract the values of $\theta_1$ and $\theta_2$ based on features in the likelihood function in \eqref{eq:maxlik}, but, even in the ideal case, these are not marked due to the weak regime. In the real case, imperfections may impart features of the same order, thus making the optimisation unable to reach meaningful values. This is supported by cases in which the variance falls below the Cram\'er-Rao limit, a clear signature of bias. The effect becomes less relevant as one reaches the intermediate condition of Fig.~\ref{fig:Kmedium} and the strong condition of Fig.~\ref{fig:Kstrong}. Due to experimental imperfections, increasing the strength would not result in considerable improvements, and the value $K=0.934$ can be considered close to the maximum achievable in the practice. In fact, $K=1$ corresponds to the limit of perfect entanglement between probe and meter.}

\color{black}{We can outline some general considerations from our experiment. The resource scaling as $K^{-2}$ should be interpreted as an optimistic consideration, as imperfections seem to have a more detrimental role as the measurement strength decreases. This can be partly remedied by collecting a larger sample. However, in the weak condition one should also expect artifacts to show up, resulting in biased estimation for which larger statistics is not a cure. The expected symmetry in the uncertainty on the two parameters is broken in the real case, with $\theta_1$ showing the largest variance. The loss of information linked to imperfection is more severe for this first parameter, accessed directly by the nondestructive measurement.}

\section{Conclusion} 
In this article we introduced and demonstrated a weak-measurement-based technique to control the sloppiness of a quantum multi-parameter estimation. We showed that, by controlling the strength of the measurement inserted between two parameters encoded in sequence, it is possible to reduce their sloppiness at the cost of increasing the intermediate system-meter interaction.  

{\color{black} These methods may be extended to multiple parameter, such as phase pairs in a multi-arm interferometer, or sequences of more than two phases in a two-arm interferometer. Our ideas of modifying the natural evolution by inserting weak measurements may still hold, but the optimal arrangement would depend on the specific setup. Our results lay groundwork in this direction, since they show how each weak measurement bears influence on the whole estimation, and should not be interpreted as an action limited to its location. }

Our results thus shed light on the interplay between sloppiness and measurement-back-action, demonstrating how an intrinsically quantum phenomenon such as weak measurements can help tackling complex multi-parameter estimation scenarios in an innovative way.


{\it Note.} During the completion of this work, Ref.~\cite{yang2025overcoming} has appeared discussing cognate methods in the context of measurements with an insufficient number of outcomes for multiparameter estimation.

{\it Funding}
This work was supported by the PRIN project PRIN22-RISQUE-2022T25TR3 of the Italian Ministry of University. G.B. is supported by Rome Technopole Innovation Ecosystem (PNRR grant M4-C2-Inv). IG acknowledges the support from MUR Dipartimento di Eccellenza 2023-2027. M.R. acknowledge support from the project PNRR - Finanziato dall’Unione eu- ropea - MISSIONE 4 COMPONENTE 2 INVESTIMENTO 1.2 - “Finanziamento di pro- getti presentati da giovani ricercatori” - Id MSCA 0000011-SQUID - CUP F83C22002390007 (Young Researchers) - Finanziato dall’Unione europea - NextGenerationEU.

\bibliography{biblio.bib}

\end{document}